\documentclass[a4paper,11pt]{article}
\usepackage{physics}
\usepackage{graphicx}  
\graphicspath{{images/}{images2/}}  
\usepackage{wrapfig} 
\usepackage{subfigure} 

\usepackage{pos}

\title{Collective effects in neutrino scattering on solid and liquid targets}

\author*[a]{Georgy Donchenko}
\author[a]{Konstantin Kouzakov}
\author[a,b]{Alexander Studenikin}

\affiliation[a]{Faculty of Physics, Lomonosov Moscow State University\\ Moscow 119991, Russia}

\affiliation[b]{Joint Institute for Nuclear Research,\\
Dubna 141980, Moscow Region, Russia}

\emailAdd{dongosha@gmail.com}
\emailAdd{kouzakov@gmail.com}
\emailAdd{studenik@srd.sinp.msu.ru}

\abstract{Neutrino scattering on liquid and solid targets at low energy transfer can serve both as a tool for searching the physics beyond the Standard Model, for example, such as neutrino electromagnetic interactions, and as a test of the Standard Model at low-energy scale. At the same time, the theoretical apparatus for low-energy elastic neutrino scattering on electrons and nuclei bound in liquids and solids must take into account collective effects in the electron and nuclear subsystems of the target. We develop such an apparatus based on the formalism of the dynamic structure factor. A numerical illustration of the role of the collective effects is provided for the case of a superfluid $^4$He target.
}

\FullConference{%
  *** The European Physical Society Conference on High Energy Physics (EPS-HEP2021), ***\\
  *** 26-30 July 2021 ***\\
  *** Online conference, jointly organized by Universität Hamburg and the research center DESY ***
}

\begin{document}
\maketitle

\section{Introduction}
    At present, detectors to search for light particles of dark matter are being discussed. In order to achieve the sensitivity for low-energy signals at the level of $\sim$1 meV, condensed matter targets are proposed \cite{Maris2017}. Such detectors can also be used to study low-energy neutrino scattering aiming both to test the Standard Model and to search for the physics beyond the Standard Model \cite{Cadeddu}. Below we show how the collective effects in neutrino scattering on a condensed matter system should be taken into account and outline the role of these effects in low-energy neutrino scattering on a superfluid $^4$He target.

\section{Elastic neutrino-atom scattering}
      In the process of elastic neutrino scattering on an atom the latter recoils as a whole and its internal state remains unchanged. We will assume that the neutrino energy satisfies the conditions $ E_\nu \ll m $ and $ E_\nu \ll 1 / R_{\rm nucl} $, where $ m $ is the atomic mass and $ R_{\rm nucl} $ is the radius of the atomic nucleus. The kinetic energy of the recoil atom is equal to the energy transfer
        \begin{equation}\label{eq: relations}
        T\leq\frac{2E_\nu^2}{m}\ll E_\nu.
        \end{equation}
		%

        The differential cross section for elastic neutrino-atom scattering is given by the following expression~\cite{Gaponov}:	        %
        \begin{equation}
					\label{cr_sec_atom}
                      \left( \frac{d\sigma}{dT} \right)_{\rm atom} = \frac{G_F^2 m}{\pi}  \left[ C_V^2 \left(1 - \frac{mT}{2E_\nu^2} \right) + C_A^2 \left(1 + \frac{mT}{2E_\nu^2} \right) \right].
        \end{equation}
        Here
		\begin{align}
							\label{C_V}
            		    C_V &= Z \left(\frac{1}{2} - 2 \sin^2 \theta_W \right) - \frac{1}{2} N + Z \left( \pm\frac{1}{2} + 2 \sin^2 \theta_W \right) F_{\rm el} ( q^2 ),\\
            	\label{C_A}	    
    					C_A^2 &= \frac{g_A^2}{4} \left[ (Z_{+} - Z_{-} ) - (N_{+} - N_{-}) \right]^2 + \frac{1}{4}\sum_{n=1,2,\dots}\sum_{l=0}^{n-1}\left|\left(  L^{nl}_{+} - L^{nl}_{-} \right) F_{\rm el}^{nl}(q^2)\right|^2,
            \end{align}
        where $q$ is the momentum transfer, with $q^2=2mT$, the plus (minus) stands for $\nu_e$ and $\bar{\nu}_e$ ($\nu_{\mu,\tau}$ and $\bar{\nu}_{\mu,\tau}$), and 
        $Z$ ($N$) is the number of protons (neutrons). $F_{\rm el}(q^2)$ ($F_{\rm el}^{nl}(q^2)$) is the Fourier transform of the electron density (the electron density in the $nl$ atomic orbital), $g_A=1.25$, and $Z_\pm$ and $N_\pm$ ($L^{nl}_\pm$) are the numbers of protons and neutrons (electrons) with spin parallel ($+$) or antiparallel ($-$) to the nucleus spin (the total electron spin).


%
\section{Neutrino scattering on a system of atoms}
    Consider now low-energy neutrino scattering by the system of  $ \mathcal{N} $ interacting atoms (liquid or solid). The energy $ T $ transferred by the neutrino to one of the atoms can be redistributed between the atoms in the system due to their interaction. The initial (final) state of the system and its energy are $ | i \rangle $ ($ | f \rangle $) and $ E_i $ ($ E_f $). For a single-atom system ($ \mathcal{N} = 1 $) we have
    \begin{equation}
    \label{d3s_if}
    \frac{d\sigma_{i\to f}}{dT d{q}^2d{\varphi_q}}=\frac{G_F^2 }{4\pi^2}  \left[ C_V^2 \left(1 - \frac{{q}^2}{E_\nu^2} \right) + C_A^2 \left(1 + \frac{{q}^2}{E_\nu^2} \right) \right]\delta(T-E_f+E_i),
    \end{equation}
	where $ \varphi_q $ is the azimuthal angle of the momentum transfer ${\vec q} $ (the $ z $ axis is directed along the initial neutrino momentum), $ E_f-E_i ={q^2} / 2m $. The result~(\ref{d3s_if}) is generelized to the case of $ \mathcal{N} $ atoms by means of the following substitutions:
    \begin{equation}
    \label{C_N}
    C_V^2\to\left|C_V^{(\mathcal{N})}\right|^2=\left|\langle f|\sum_{j=1}^{\mathcal{N}}e^{i\vec{q}\vec{R}_j}C_V|i\rangle\right|^2, \qquad C_A^2\to\left|C_A^{(\mathcal{N})}\right|^2=\left|\langle f|\sum_{j=1}^{\mathcal{N}}e^{i\vec{q}\vec{R}_j}C_A|i\rangle\right|^2,
    \end{equation}
    where $ \vec{R}_j $ denotes the position of the $j$th atom. 
	Summing over all possible final states and averaging over the initial states, we find
	\begin{align}
	\label{cr_sec}
	\frac{d\sigma}{dT}=&\int\limits_0^{\infty} d{q}^2 \int\limits_0^{2\pi} d{\varphi_q}\,\frac{G_F^2 }{4\pi^2}  \left[ C_V^2 \left(1 - \frac{{q}^2}{E_\nu^2} \right) + C_A^2 \left(1 + \frac{{q}^2}{E_\nu^2} \right) \right]S(T,{\vec q}).
	\end{align}
	Here we introduced the dynamic structure factor~\cite{Kittel,Kvasnikov}
    \begin{equation}
						\label{S(T,q)}
        S(T, \vec{q}) = \sum_{i,f} w_i \left|\langle f|\sum_{j=1}^{\mathcal{N}}e^{i\vec{q}\vec{R}_j}|i\rangle\right|^2\delta(T-E_f+E_i),
    \end{equation}
	where $ w_i $ is the statistical weight of the state $ | i \rangle $. 

\section{Neutrino scattering on superfluid $^4$He}
   We consider neutrino scattering by the liquid $^4$He in the superfluid phase to illustrate the developed formalism and to point out the role of the collective effects in the target.
    The dynamic structure factor in the discussed range of the energy-transfer values ($\lesssim1$ meV) can be approximated as $ S(T, \vec{q}) = \langle\rho_{\vec q}\rho_{\vec q}^+\rangle\delta(T-uq)$~\cite{Kvasnikov}, where $u$ is the sound velocity in superfluid $^4$He. 
    For the cross section~(\ref{cr_sec}) we obtain
	\begin{align}
	\label{cr_sec_phonon}
	\frac{d\sigma_0}{dT}=\mathcal{N}C_V^2\,\frac{G_F^2 T}{2 \pi m u^2 } \left(1 - \frac{T^2}{4u^2E_\nu^2} \right).
	\end{align}
	    At the same time, for the system of  $ \mathcal{N} $ non-interacting helium atoms we have        
    \begin{equation}
				\label{cr_sec_atom_He}
                  \frac{d\sigma}{dT} = \mathcal{N}C_V^2\,\frac{G_F^2 m}{\pi}\left(1 - \frac{mT}{2E_\nu^2} \right).
    \end{equation}

\begin{figure}[t]
        \includegraphics[width=0.8\linewidth]{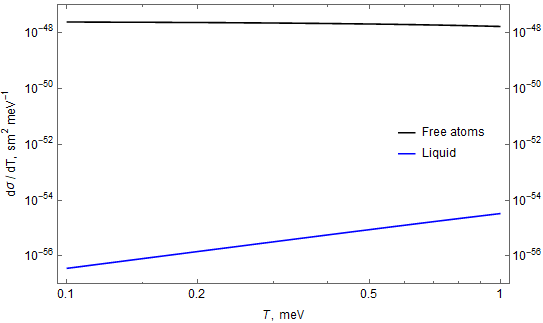}
               \centering
        \caption{The differential cross section for tritium antineutrino scattering on the system of helium atoms at $E_\nu=10$~keV normalized to the number of atoms $\mathcal{N}$.} \label{fig:cs}
        \centering
    \end{figure}

    The qualitative difference between Eqs.~(\ref{cr_sec_phonon}) and~(\ref{cr_sec_atom_He}) is obvious. Fig.~\ref{fig:cs} shows the cross section of tritium antineutrino scattering on the superfluid $^4$He in comparison with that on free helium atoms. The numerical calculations show that the cross section with account for collective effects is strongly (by a factor of $10^6$) suppressed relative to the case of free atoms. 
    
\section{Conclusions}
        The theory of low-energy neutrino scattering on a condensed-matter system has been developed. It has been shown that taking collective effects into account in the neutrino scattering by the superfluid $^4$He qualitatively changes the dependence of the differential cross section on the energy transfer. This fact must be taken into account both in the preparation and in the data analysis of future neutrino experiments with detectors based on the liquid $^4$He and other materials (for example, such as graphene~\cite{Hochberg2017}). The obtained results can be used in the search for the electromagnetic properties of neutrinos~\cite{RMP2015}. 

\vspace{\baselineskip}
This research has been supported by the Interdisciplinary Scientific and Educational School of Moscow University ``Fundamental and Applied Space Research'' and also by the Russian Foundation for Basic Research under Grant No. 20-52-53022-GFEN-a. The work of G.D. is supported by the BASIS Foundation No. 20-2-9-9-1.

\end{document}